\documentclass[runningheads]{llncs}
\usepackage{graphicx}
\usepackage[nolist]{acronym}
\usepackage{hyperref}
\usepackage{listings}
\usepackage{verbatim}
\usepackage{listings,xcolor}
\usepackage{float}
\usepackage{tabularx,colortbl}
\usepackage{footnote}
\usepackage[caption = false]{subfig}
\usepackage{soul}
\usepackage[misc]{ifsym}
\begin{acronym}
 \acro{EDP}{European Data Portal}
 \acro{API}{Application Programming Interface}
 \acro{EC}{European Commission}
 \acro{EU}{European Union}
 \acro{ETL}{Extract, Transform, Load}
 \acro{DCAT-AP}{DCAT Application profile for data portals in Europe}
 \acro{DCAT}{Data Catalog Vocabulary}
 \acro{DQV}{Data Quality Vocabulary}
 \acro{HTTP}{Hypertext Transfer Protocol}
 \acro{RDF}{Resource Description Framework}
 \acro{MQA}{Metadata Quality Assurance}
 \acro{LOD}{Linked Open Data}
 \acro{CAS}{Central Authentication Service}
 \acro{CKAN}{Comprehensive Knowledge Archive Network}
 \acro{LDP}{Linked Data Platform}
 \acro{PDF}{Portable Document Format}
 \acro{SHACL}{Shapes Constraint Language}
 \acro{SKOS}{Simple Knowledge Organization System}
 \acro{SPARQL}{SPARQL Query Language}
 \acro{CDS}{CKAN Data Schema}
 \acro{W3C}{World Wide Web Consortium}
 \acro{XSD}{XML Schema Definition}
 \acro{XSLT}{eXtensible Stylesheet Language Transformations}
 \acro{LSH}{locality-sensitive hashing}
 \acro{SWT}{Semantic Web technologies}
 \acro{PPL}{Piveau pipeline}
\end{acronym}

\begin{document}
\title{Piveau: A Large-scale Open Data Management Platform based on Semantic Web Technologies}
\titlerunning{Piveau}
%
%
\author{Fabian Kirstein\inst{1,2} \textsuperscript{\Letter} \and
Kyriakos Stefanidis \inst{1}
Benjamin Dittwald\inst{1} \and
Simon Dutkowski\inst{1}
\and Sebastian Urbanek \inst{1,2}
\and Manfred Hauswirth\inst{1,2,3}
} 
\authorrunning{F. Kirstein et al.}
\institute{Fraunhofer FOKUS, Berlin, Germany \and
Weizenbaum Institute for the Networked Society, Berlin, Germany \and TU Berlin, Open Distributed Systems, Berlin, Germany
\\
\email{\{firstname.lastname\}@fokus.fraunhofer.de}}
\maketitle              
\begin{abstract}
The publication and (re)utilization of Open Data is still facing multiple barriers on technical, organizational and legal levels. This includes limitations in interfaces, search capabilities, provision of quality information and the lack of definite standards and implementation guidelines. Many Semantic Web specifications and technologies are specifically designed to address the publication of data on the web. In addition, many official publication bodies encourage and foster the development of Open Data standards based on Semantic Web principles.  However, no existing solution for managing Open Data takes full advantage of these possibilities and benefits. In this paper, we present our solution “Piveau”, a fully-fledged Open Data management solution, based on  Semantic Web technologies. It harnesses a variety of standards, like RDF, DCAT, DQV, and SKOS, to overcome the barriers in Open Data publication. The solution puts a strong focus on assuring data quality and scalability. We give a detailed description of the underlying, highly scalable, service-oriented architecture, how we integrated the aforementioned standards, and used a triplestore as our primary database. We have evaluated our work in a comprehensive feature comparison to established solutions and through a practical application in a production environment, the European Data Portal. Our solution is available as Open Source.

\end{abstract}

\section{Introduction}
Open Data constitutes a prospering and continuously evolving concept. At the very core, this includes the publication and re-utilization of datasets. Typical actors and publishers are public administrations, research institutes, and non-profit organizations. Common users are data journalists, businesses, and governments. The established method of distributing Open Data is via a web platform that is responsible for gathering, storing, and publishing the data. Several software solutions and specifications exist for implementing such platforms. Especially the \ac{RDF} data model and its associated vocabularies represent a foundation for fostering interoperability and harmonization of different data sources. The \ac{DCAT} is applied as a comprehensive model and standard for describing datasets and data services on Open Data platforms \cite{noauthor_data_nodate}. However, \ac{RDF} is only a subset of the Semantic Web stack and Open Data publishing does not benefit from the stack's full potential, which offers more features beyond  data modeling.  Therefore, we developed a novel and scalable platform for managing Open Data, where the Semantic Web stack is a first-class citizen. Our work focuses on two central aspects: (1) The utilization of a variety of Semantic Web standards and technologies for covering the entire life-cycle of the Open Data publishing process. This covers particularly data models for metadata, quality verification, reporting, harmonization, and machine-readable interfaces. (2) The application of state-of-the-art software engineering approaches for development and deployment to ensure production-grade applicability and scalability. Hence, we integrated a tailored microservice-based architecture and a suitable orchestration pattern to fit the requirements in an Open Data platform. \\
It is important to note, that currently our work emphasizes the management of metadata, as intended by the \ac{DCAT} specification. Hence, throughout the paper the notion of data is used in terms of metadata. \\
In Section 2 we describe the overall problem and in Section 3 we discuss related and existing solutions. Our software architecture and orchestration approach is described in Section 4. Section 5 gives a detailed overview of the data workflow and the applied Semantic Web standards. We evaluate our work in Section 6 with a feature analysis and an extensive use case. To conclude, we summarize our work and give an outlook for future developments.

\section{Problem Statement}\label{sec:problem}
A wide adoption of Open Data by data providers and data users is still facing many barriers. Beno et al. \cite{beno_perception_2017} conducted a comprehensive study of these barriers, considering legal, organizational, technical, strategic, and usability aspects. Major technical issues for users are the limitations in the \acp{API}, difficulties in searching and browsing, missing information about data quality, and language barriers. Generally, low data quality is also a fundamental issue, especially because (meta)data is not machine-readable or, in many cases, incomplete. In addition, low responsiveness and bad performance of the portals have a negative impact on the adoption of Open Data. For publishers, securing the integrity and authenticity, enabling resource-efficient provision, and clear licensing are highly important issues. The lack of a definite standard and technical solutions is listed as a core barrier. \\
The hypothesis of our work is, that \textbf{a more sophisticated application of Semantic Web technologies can lower many barriers in Open Data publishing and reuse}. These technologies intrinsically offer many aspects, which are required to improve the current support of Open Data. Essentially, the Semantic Web is about defining a common standard for integrating and harnessing data from heterogeneous sources \cite{W3CSemanticWeb}. Thus, it constitutes an excellent match for the decentralized and heterogeneous nature of Open Data.\\
Widespread solutions for implementing Open Data platforms are based on canonical software stacks for web applications with relational and/or document databases. The most popular example is the Open Source solution \ac{CKAN} \cite{noauthor_ckan_nodate}, which is based on a flat JSON data schema, stored in a PostgreSQL database. This impedes a full adoption of Semantic Web principles. The expressiveness of such a data model is limited and not suited for a straightforward integration of \ac{RDF}. 

\section{Related Work}\label{sec:related}

Making Open Data and Linked Data publicly available and accessible is an ongoing process that involves innovation and standardization efforts in various topics such as semantic interoperability, data and metadata quality, standardization as well as toolchain and platform development. 

One of the most widely adopted standards for the description of datasets is DCAT and its extension \ac{DCAT-AP} \cite{european_commission_about_2019}. The latter adds metadata fields and mandatory property ranges, making it suitable for use with Open Data management platforms. Its adoption by various European countries led to the development of country-specific extensions such as the official exchange standard for open governmental data in Germany \cite{noauthor_dcat-ap.-_nodate} and Belgium's extension \cite{dcatbe}. Regarding Open Data management platforms, the most widely known Open Source solution is \ac{CKAN} \cite{noauthor_ckan_nodate}. It is considered the de-facto standard for the public sector and is also used by private organizations. It does not provide native Linked Data capabilities but only a mapping between existing data structures and RDF. Another widely adopted platform is uData \cite{uData}. It is a catalog application for collecting data and metadata focused on being more contributive and inclusive than other Open Data platforms by providing additional functionality for data reuse and community contributions. Other Open Source alternatives include the repository solution DSpace which dynamically translates \cite{noauthor_linked_nodate-1} relational metadata into native RDF metadata and offers it via a SPARQL endpoint. WikiData also follows a similar approach \cite{vrandecic_wikidata:_2014}; it uses a custom structure for identifiable items, converts them to native RDF and provides an API endpoint. Another, proprietary, solution is OpenDataSoft \cite{noauthor_opendatasoft_nodate}, which has limited support for Linked Data via its interoperability mode. There are also solutions that offer native Linked Data support following the W3C recommendation for Linked Data Platforms (LDPs). Apache Marmotta \cite{noauthor_ldp_nodate} has native implementation of RDF with a pluggable triplestore for Linked Data publication. Virtuoso \cite{noauthor_openlink_nodate} is a highly scalable LDP implementation that supports a wide array of data access standards and output formats. Fedora \cite{fedora} is a native Linked Data repository suited for digital libraries. Recent research efforts \cite{Roffia2018} focusses on the notion of dynamic Linked Data where context aware services and applications are able to detect changes in data by means of publish-subscribe mechanisms using SPARQL. 

A core feature of most big commercial platforms is the \ac{ETL} functionality. It refers to the three basic data processing stages of reading data (extract) from heterogeneous sources, converting it (transform) to a suitable format, and storing it (load) into a database. Platforms that offer ETL as a core functionality include IBM InfoSphere \cite{IBMInfosphere} with its DataStage module, Oracle Autonomus Data Warehouse \cite{OracleADW} with its Data Integrator module and SAS Institute's data warehouse \cite{SASInstitute}. Moreover, various Open Source solutions such as Scriptella \cite{Scriptella} and Talend Open Studio \cite{Talend} are based on ETL. The above data warehouses offer highly scalable ETL functionality but do not support Linked Data and DCAT. On the other hand, the previously mentioned Linked Data platforms do not offer any real ETL capabilities. Bridging this gap was the main objective that led to the development of the Piveau pipeline as a core part of our architecture. Similar data pipelines can be found as stand-alone services and applications such as AWS Data Pipeline \cite{AWSDataPipeline}, Data Pipes from OKFN \cite{OKFLDatapipes}, North Concepts Data Pipeline \cite{NCDataPipeline}, and Apache Airflow \cite{ApacheAirflow}.

\section{A Flexible Architecture for Semantic Web Applications}
Semantic Web technologies are mainly supported by specifications, standards, libraries, full frameworks, and software. The underlying concept of our architecture is the encapsulation of Semantic Web functionalities to make them reusable and interoperable, which is considered a classical software engineering principle. Our Open Data platform introduces a state-of-the-art, tailored architecture to orchestrate these encapsulations and make them easy to apply in production environments. It is based on a microservice architecture and a custom pipeline system, facilitating a flexible and scalable feature composition of Open Data platforms. This enables the application of Piveau for various use cases and audiences. Furthermore, it enables the re-use of features in other environments and applications.

\subsection{The Piveau Pipeline}\label{sec:pipe}
The basic requirements of our architecture were the use of microservices, high scalability, lightweight in application and management, and suitable for large-scale data processing. Existing workflow engines and \ac{ETL} systems are either not designed for Linked Data and/or limited solely to extensive data integration tasks (see Section \ref{sec:related}). To lower complexity and maintenance needs, we aimed for an unifying architecture and data processing concept, which targets specifically our needs. Therefore, we designed and implemented the \ac{PPL}. The \ac{PPL} builds upon three principal design choices: (1) All services and features expose RESTful interfaces and comply with the microservice style. (2) The services can be connected and orchestrated in a generic fashion to implement specific data processing chains. (3) There is no central instance, which is responsible for orchestrating the services. 

A \ac{PPL} orchestration is described by a \emph{descriptor}, which is a plain JSON document, including a list of segments, where each segment describes a step (a service) in the data processing chain. Every segment includes at least meta-information, targeting the respective service and defining the consecutive service(s).\footnote{The PPL descriptor schema can be found at: \url{https://gitlab.com/piveau/pipeline/piveau-pipe-model/-/blob/master/src/main/resources/piveau-pipe.schema.json}} The entire descriptor is passed from service to service as state information. Each service identifies its segment by a service identifier, executes its defined task and passes the descriptor to the next service(s). Hence, the descriptor is a compilation and self-contained description of a data processing chain. Each microservice must expose an endpoint to receive the descriptor and must be able to parse and execute its content. The processed data itself can be embedded directly into the descriptor or passed via a pointer to a separate data store, e.g. a database, file system or other storage. This depends on the requirements and size of data and can be mixed within the process.

The \ac{PPL} has been proven to be a fitting middle ground between \ac{ETL} approaches and workflow engines. On an architectural level, it allows to harvest data from diverse data providers and orchestrate a multitude of services. Its production-level implementation in the \ac{EDP} supports millions of open datasets with tens of thousands updates per day (see Section 6.2).

\subsection{Architecture, Stack and Deployment}

\begin{figure}
   {\includegraphics[width=\textwidth]{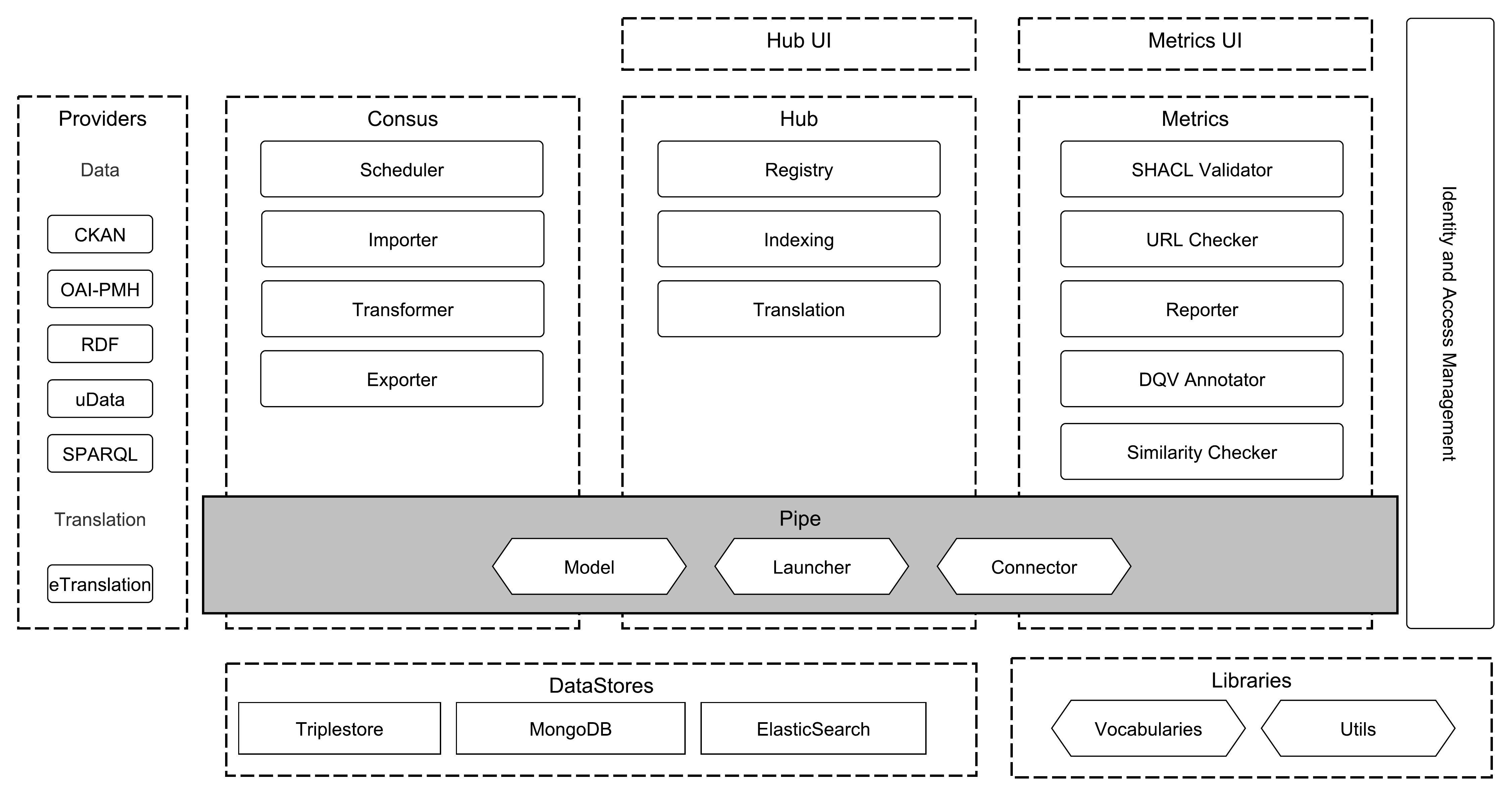}}
\caption{Piveau High-Level Architecture} \label{fig:overview}
\end{figure}

The development of Piveau follows the reactive manifesto, which requires a system to be responsive, resilient, elastic, and message driven \cite{reactive_manifesto}. The platform is divided into three logical main components, each one responsible for a phase within the life-cycle of the datasets: Consus, Hub and Metrics. Figure \ref{fig:overview} illustrates the overall architecture and structure. 

Consus is responsible for the data acquisition from various sources and data providers. This includes scheduling, transformation and harmonization. Hub is the central component to store and register the data. Its persistence layer consists of a Virtuoso triplestore\footnote{\url{https://virtuoso.openlinksw.com/}} as the principal database, Elasticsearch\footnote{\url{https://www.elastic.co/products/elasticsearch}} as the indexing server and a MongoDB\footnote{\url{https://www.mongodb.com/}} for storing binary files. Metrics is responsible for creating and maintaining comprehensive quality information and feeding them back to the Hub. Two web applications based on Vue.js\footnote{\url{https://vuejs.org/}} are available for browsing the data. The services are written with the reactive JVM framework Vert.x\footnote{\url{https://vertx.io/}} and orchestrated with the \ac{PPL} within and across the logical components. Several libraries for common tasks, \ac{RDF} handling and the \ac{PPL} orchestration are re-used in all services.\\  
In order to enable native cloud deployment, we use the Docker\footnote{\url{https://www.docker.com/}} container technology. Each service is packaged as a container, supporting easy and scalable deployment. In addition, Piveau was tested with Kubernetes-based\footnote{\url{https://kubernetes.io/}} container management solutions like Rancher\footnote{\url{https://rancher.com/}} and OpenShift\footnote{\url{https://www.openshift.com/}}. Hence, our architecture supports a production-grade development scheme and is ready for DevOps practices. 

\subsection{Security Architecture}\label{sec:security}
In this section we will describe how Piveau handles authentication, authorization, and identity management. The multitude of standardized system and network security aspects that are part of the Piveau architectural design, such as communication encryption, firewall zones and API design, are beyond the scope of this paper.

Piveau is comprised of multiple microservices, Open Source software and a set of distinct web-based user interfaces. In order to support Single Sign-On (SSO) for all user interfaces and authentication/authorization to all microservices, we use Keycloak\footnote{\url{https://www.keycloak.org/}} as central identity and access management service. Keycloak also supports federated identities from external providers. Specifically, in the case of the \ac{EDP}, we use "EU Login" as the sole external identity provider without allowing any internal users apart from the administrators.
Authentication and authorization on both front-end and back-end services follows the OIDC protocol \cite{OIDC1.0}. More specifically, all web-based user interfaces follow the OIDC authorization code flow. This means that when a user tries to login to any of Piveau's user interfaces, they are redirected to the central Keycloak authentication form (or the main identity provider's authentication form) and, upon successful login, they are redirected back to the requested web page. This provides a uniform user experience and minimizes the risk of insecure implementation of custom login forms.

All back-end services also follow OIDC by requiring valid access tokens for each API call. Those tokens follow the JSON Web Token (JWT) standard. In contrast to static internal API keys, this design pattern supports arbitrary back-end services to be open to the public without any change to their authentication mechanisms. Moreover, since the JWT tokens are self-contained, i.e. they contain all the required information for user authentication and resource authorization, the back-end services can perform the required checks without the need of communication with a database or Keycloak. Not requiring round-trips greatly enhances the performance of the whole platform.

The fine-grained authorization follows the User-Managed Access (UMA) specification \cite{uma2.0}, where resource servers (back-end services) and a UMA-enabled authorization server (Keycloak) can provide uniform management features to user-owned resources such as catalogs and datasets.

\section{Semantic Data Workflow}\label{sec:workflow}

\begin{figure}
\centering
   {\includegraphics[width=0.85\textwidth]{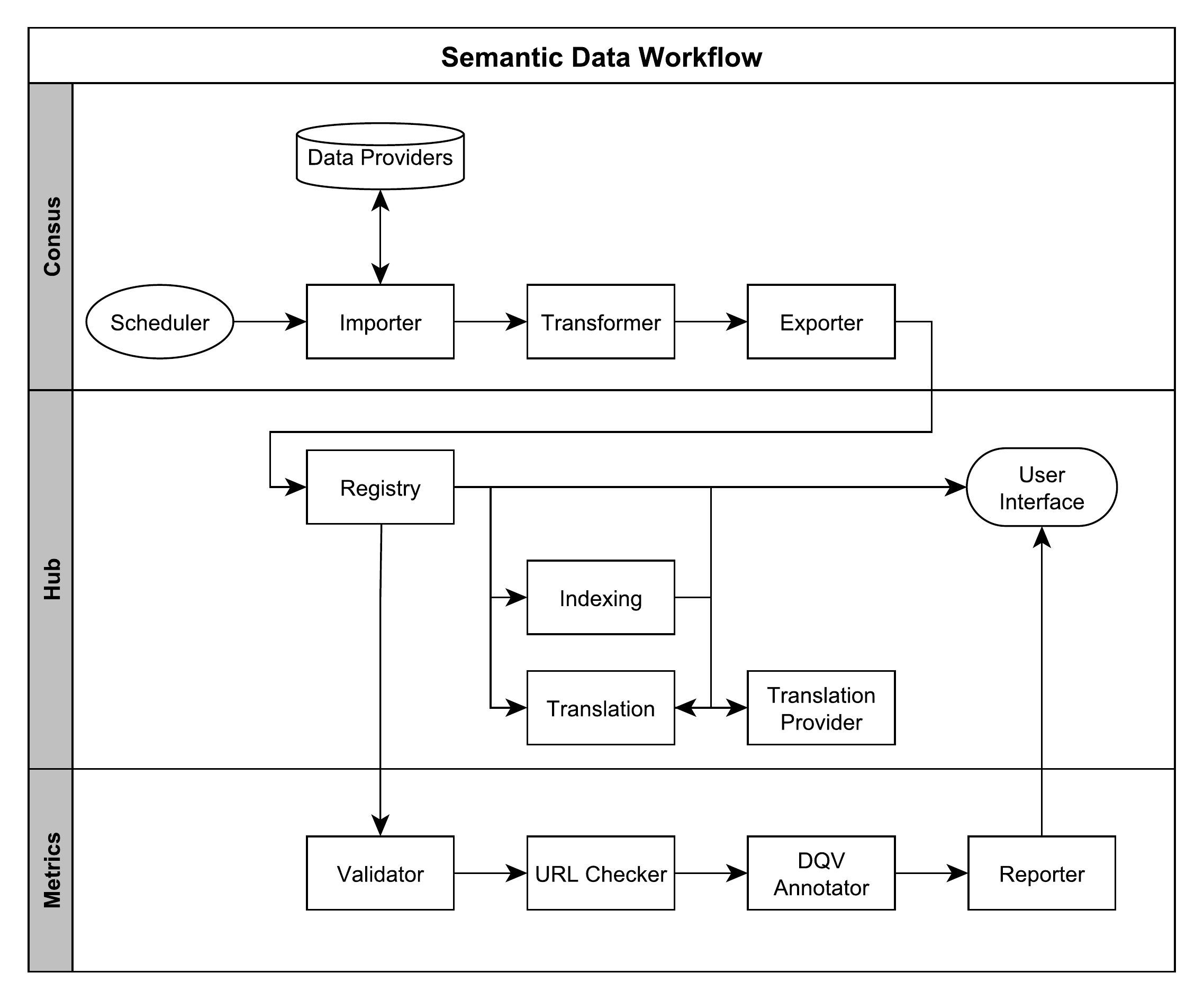}}
\caption{Semantic Data Workflow} \label{fig:workflow}
\end{figure}

In the following, a typical data flow in our Open Data platform is described to illustrate our solution in detail. This covers the process of acquiring the data from the original providers, evaluating the quality of that data, and presenting and managing the data (see Figure \ref{fig:workflow}). We focus on the used Semantic Web technologies and specifications. The presented order reflects roughly the order of execution. But since many processes run asynchronously, the order can vary depending on their execution time. 

\subsection{Data Acquisition}

The main entry point for any data workflow and orchestration is the \textbf{scheduler}. Each data workflow, defined as a \ac{PPL} descriptor (see Section \ref{sec:pipe}), is assigned a list of triggers. A trigger may define a periodical execution (hourly, daily, weekly, bi-weekly, yearly, etc.), number of execution times, a list of specific date and times to execute, or an immediate execution. Each trigger is able to pass its own process configuration in order to individualize the workflow depending on the execution time. Upon execution, the scheduler passes the descriptor to the first service in line, typically an \textbf{importer}. \\
An importer retrieves the metadata from the source portal(s). We have implemented a range of importers to support a variety of interfaces and data formats, e.g. CKAN-API, OAI-PMH, uData, RDF, and SPARQL. The importer is responsible for extracting records of metadata from either an API or a dump file and for sending it to the next processing step. This covers the generation of a complete list of identifiers of all datasets, which will be required for a final synchronization, including the deletion of datasets, which are not present in the source portal anymore.\\
The principal data format of Piveau is \ac{RDF}, therefore non-\ac{RDF} or not supported \ac{RDF} dialects sources require a transformation. A \textbf{transformer} generates \ac{RDF} from such source data, by applying light-weight transformation scripts written in JavaScript. The final output is always DCAT-compliant RDF. The scripts can be managed externally (e.g. in Git) to ensure maintainability. \\
Finally, our \textbf{exporter} sends the \ac{RDF} data to the Hub component. Non-existing datasets are deleted by the exporter based on the identifier list that is acquired in the importing step. 

\subsection{Processing and Storing}
The central service for dataset management is the \textbf{registry}. It acts as a middleware and abstraction layer to interact with the triplestore. It offers a RESTful interface, supporting the major \ac{RDF} serializations (Turtle, JSON-LD, N-Triples, RDF/XML, Notation3). Its resources reflect the main \ac{DCAT} entities: catalog, dataset, and distribution. The main task is to pre-process and harmonize the data received from the exporter. This includes the application of consistent and meaningful URI schemata \cite{noauthor_10_nodate}, the generation of unique IDs, and the mapping to linked, existing entities. It ensures the integrity and traceability of the data in the triplestore. 
The \textbf{indexing} service is responsible for managing the high-performance search index. It receives the processed \ac{RDF} data from the registry and flattens it into a plain JSON representation, which is suitable for indexing. Firstly, this is done by extracting relevant literals from the data, e.g. from properties like title and description. Secondly, linked resources are resolved and proper literals are extracted from the result (for instance by looking for \emph{rdfs:label}). The service supports the use of existing and well-maintained vocabularies and ontologies for that purpose. Piveau ships with a selection of vocabularies, e.g. for human languages, licenses, and geolocations. The result of the search service constitutes one of the main access points to the data, because it is much more human-readable than native \ac{RDF}. \\
The \textbf{translation} service manages the machine translation of literals into multiple languages. It represents a middleware to third-party translations services, bundling strings from multiple datasets to an integrated request. After completion the service stores the translation by applying the native multi-language features of RDF. As soon as a dataset is retrieved, the existing original languages are identified and added to the text information using a language tag inside the dataset. This labeling is based on ISO 639-1 language codes. In addition, metadata about the translation status are stored in the dataset, indicating when a translation was started and when it was completed. Translated text information are labeled with an extended language tag to differentiate them from the original text. It follows the schema \emph{en-t-de-t0-abc} \cite{rfc_6497}, where the target language is named first, followed by a \emph{t} and the original language. \\
Finally, the data is accessible via multiple means. The triplestore exposes a SPARQL endpoint, which offers raw und direct access to the data. A RESTful \ac{API} allows the access to the
\ac{RDF} serializations, provided by the registry and to the indexed serializations, provided by the search service. A web \textbf{user interface} offers access to end users and interacts directly with the RESTful \ac{API}.

\subsection{Quality Evaluation}
In parallel with the main data processing steps, the data is processed by dedicated services to assess its quality. Semantic Web technologies offer mature tools and standards to conduct this task. \\
The \textbf{validator} provides a formal validation of each dataset. We apply the W3C \ac{SHACL} \cite{SHACL}, where a pre-defined set of rules is tested against a dataset. Currently the DCAT-AP \ac{SHACL} rules \cite{dcat-ap__shacl_121} are included. The validation results include detailed information about issues and violations. This result covers the exact paths and reasons for the identified deficits. The applied rules can also be extended or replaced. In addition, the \textbf{URL checker} performs accessibility tests on each linked distribution (the actual data) and assesses its availability via HTTP status codes. \\
The \textbf{DQV annotator} \cite{DQL} provides a qualitative assessment for each dataset. It is based on a custom metrics scheme, which is inspired by the FAIR principles \cite{fair}. The findability dimension refers to completeness of the metadata, e.g. whether keywords, geo data or time information are provided. Accessibility refers to the results from the URL checker. Interoperability is assessed by evaluating the format and type of data, which is referenced in a dataset (distribution). For instance, if the data is in a machine-readable and/or non-proprietary format. Reusability is mostly confirmed by checking the availability of licensing information. Beyond this FAIR evaluation, the similarity of a dataset to other datasets is calculated based on \ac{LSH} algorithm. \\
The results of the validation and annotator services are summarized in a quality report and attached as \ac{RDF} to the concerned dataset in the triplestore. This report uses a custom quality vocabulary, which applies the W3C \ac{DQV} and reflects our metric scheme. In addition, an aggregated report is attached to the respective catalog. \\
The \textbf{reporter} offers a variety of human-readable versions of the quality reports. It collects all data from the triplestore and renders visually appealing reports of the information. It supports PDF, XLS or ODS. In addition, a comprehensive web front-end is available, and is integrated into the front-end of the Hub component.        

\section{Evaluation}
We have evaluated our work according to three quantitative and qualitative aspects. In Section \ref{sec:features} we compare Piveau with two well-known Open Data solutions. In Section \ref{sec:edp} we describe a real-world application based on Piveau. Finally, in Section \ref{sec:swtech} we present an analysis of the impact of Semantic Web technologies on the perceived barriers of Open Data.

\subsection{Feature Comparison with Open Data Solutions}\label{sec:features}
No definite metric exists to specifically assess the technical performance of Open Data technologies and infrastructures. However, a lot of work and research was conducted in the field of requirements and evaluation modeling for Open Data. An extensive review covering a broad variety of dimensions (economical, organizational, ergonomic, etc.) is presented by Charalabidis et al. \cite{WorldOpenData2018} This includes an overview of "Functional Requirements of an Open Data Infrastructure", which acts as the main basis for our feature matrix \cite{WorldOpenData2018}. It is supplemented by indicators from the outcome of "Adapting IS [Information Systems] Success Model on Open Data Evaluation" \cite{WorldOpenData2018}. Furthermore, we translated the W3C recommendation for best practices for publishing data on the web into additional indicators \cite{w3c_data_on_the_web_bp}. Finally, the matrix is complemented by custom indicators to reflect our experiences in designing and developing Open Data infrastructures. In the selection process we only focused on indicators, which were applicable to measurable technical aspects that reflect the overall objective of managing metadata. More personal indicators, like "The web pages look attractive", were not considered. Still, this approach led to a large number of indicators (\textgreater50), which we semantically combined to generate a compact and meaningful feature matrix.\footnote{The exact provenance and creation process of the feature matrix is available as supplementary material:  \url{https://zenodo.org/record/3571171}}\\
We compared Piveau with the popular Open Data solutions CKAN and uData (see Section \ref{sec:related}). The selection criteria were: (1) Must be freely available as Open Source software; (2) Must not be a cloud- or hosting-only solution; (3) Has a high rate of adoption and (4) Primarily targets public sector data. Table 1 shows the final feature matrix and the result of the evaluation. Each measure was rated with the following scale: 0 - not supported, 1 - partially supported, 2 - fully supported. An explanation is given for each rating, where required.

\begin{table}
   \begin{tabular}
   {>{\raggedright\tiny}p{4.0cm}|>{\tiny}c|>{\raggedright\tiny}p{2.37cm}|>{\tiny}c|>{\raggedright\tiny}p{2.37cm}|>{\tiny}c|>{\tiny}p{2.37cm}}
   \multicolumn{1}{l}{}& \multicolumn{2}{c}{\scriptsize{Piveau}} & \multicolumn{2}{c}{\scriptsize{CKAN}} & \multicolumn{2}{c}{\scriptsize{uData}} \\
   \hline\hline

   \multicolumn{4}{l}{\scriptsize{\textbf{Searching and Finding Data}}} \\
   \hline
   Support for data federation & 2 & \emph{Native support through SPARQL} & 1 & \emph{Indirect through harvesting} & 1 & \emph{Indirect through harvesting} \\
   \hline
   Integration of controlled vocabularies & 2 & \emph{Support for structured controlled vocabulary} & 1 & \emph{Support for simple controlled vocabulary} & 1 & \emph{Support for simple controlled vocabulary} \\
   \hline
   Filtering, sorting, structuring, browsing and ordering search results by diverse dimensions & 2 & \emph{Application of search engine} & 2 & \emph{Application of search engine} & 2 & \emph{Application of search engine} \\
   \hline
   Offer a strong and interoperable API & 2 & \emph{DCAT compliant REST} & 2 & \emph{DCAT compliant REST} & 2 & \emph{DCAT compliant REST} \\
   \hline
   Support multiple languages & 2 & \emph{On interface and dataset level} & 1 & \emph{Only on interface level} & 2 & \emph{On interface and dataset level} \\
   \hline
   Linked Data interface & 2 & \emph{SPARQL endpoint} & 0 & \emph{} & 0 & \emph{} \\
   \hline
   Geo-Search & 2 & \emph{Available} & 2 & \emph{Available} & 2 & \emph{Available} \\
   \hline \hline

   \multicolumn{4}{l}{\scriptsize{\textbf{Data Provision and Processing}}} \\
   \hline
   Data Upload & 1 & \emph{Binary data upload} & 2 & \emph{Binary and structured data upload} & 1 & \emph{Binary data upload} \\
   \hline
   Data Enrichment and Cleansing & 0 & \emph{} & 0 & \emph{} & 0 & \emph{} \\
   \hline
   Support for linking and referring other data & 2 & \emph{Any number of links possible} & 1 & \emph{Restrictive schema} & 1 & \emph{Restrictive schema} \\
   \hline \hline

   \multicolumn{4}{l}{\scriptsize{\textbf{Analysis and Visualization}}} \\
   \hline
   Provide comprehensive metadata & 2 & \emph{Complete and extensible schema} & 1 & \emph{Restricted schema} & 1 & \emph{Restricted schema} \\
   \hline
   Offer tools for analyses & 0 & \emph{} & 1 & \emph{Preview of tabular data} & 0 & \emph{} \\
   \hline
   Visualizing data on maps & 1 & \emph{Visualization of geo metadata} & 1 & \emph{Visualization of geo metadata} & 1 & \emph{Visualization of geo metadata} \\
   \hline
   Detailed reuse information & 0 & \emph{} & 0 & \emph{} & 1 & \emph{Indicates purpose and user} \\
   \hline \hline

   \multicolumn{4}{l}{\scriptsize{\textbf{Quality Assurance}}} \\
   \hline
   Information about data quality & 2 & \emph{Comprehensive quality evaluation} & 0 & \emph{} & 1 & \emph{Simple quality evaluation} \\
   \hline
   Provide quality dimensions to compare datasets and its evolution & 2 & \emph{Comprehensive quality evaluation} & 0 & \emph{} & 0 & \emph{} \\
   \hline \hline

   \multicolumn{4}{l}{\scriptsize{\textbf{Interaction}}} \\
   \hline
   Support interaction and communication between various stakeholders & 0 & \emph{} & 0 & \emph{} & 2 & \emph{Discussion platform} \\
   \hline
   Enrich data & 0 & \emph{} & 0 & \emph{} & 1 & \emph{Additional community resources} \\
   \hline
   Support revisions and version history & 0 & \emph{} & 1 & \emph{Metadata revision} & 0 & \emph{} \\
   \hline
   Track reuse & 0 & \emph{} & 0 & \emph{} & 2 & \emph{Linked reuse in dataset} \\
   \hline \hline

   \multicolumn{4}{l}{\scriptsize{\textbf{Performance and Architecture}}} \\
   \hline
   Maturity & 1 & \emph{Application in a few portals} & 2 & \emph{Application in many portals} & 1 & \emph{Application in a few portals} \\
   \hline
   Personalization and Custom Themes & 1 & \emph{Replaceable themes} & 2 & \emph{Use of theme API} & 1 & \emph{Replaceable themes} \\
   \hline
   Scalable Architecture & 2 & \emph{Microservice architecture} & 1 & \emph{Monolithic architecture} & 1 & \emph{Monolithic architecture} \\
   \hline\hline

   \multicolumn{1}{l}{\scriptsize{\textbf{Score}}} & \multicolumn{2}{c}{\scriptsize{28}} &  \multicolumn{2}{c}{\scriptsize{21}} & \multicolumn{2}{c}{\scriptsize{24}}
   \end{tabular}
   \caption{Feature Comparison}
   \end{table}

The overall result indicates that our solution can match with existing and established solutions and even reaches the highest score. Piveau offers strong features regarding searching and finding datasets and data provision. The comprehensive metadata is a great foundation for analyses and visualizations. Our features for quality assurance are unrivaled and we support the most scalable architecture. Yet, uData offers unique features for interaction and CKAN is very mature and industry-proven.  

\subsection{The European Data Portal}\label{sec:edp}
The \ac{EDP}\footnote{\url{https://www.europeandataportal.eu}} is a central portal, publishing all metadata of Open Data provided by public authorities of the \ac{EU}. It gathers the data from national Open Data portals and geographic information systems. It was initially launched in November 2015 by the \ac{EC}. Its design and development was driven by the DCAT-AP specification.

The \ac{EDP} was one of the first implementations of the DCAT-AP specification. In order to comply with established Open Data publishing concepts, the first version was based on an extended CKAN with an additional layer for transforming and replicating all metadata into RDF. This setup required additional mechanisms to transform data and, thus, proved to be too complex and limited for the growing amounts of Open Data in Europe. \cite{kirstein2019linked} We successfully improved this first version with our solution Piveau. This successfully enrolled our solution in a large-scale production environment. Our translation middleware integrates the eTranslation Service of the EU Commission \cite{noauthor_authority_nodate}, enabling the provision of metadata in 25 European languages. As of December 2019 the \ac{EDP} offers approximately one million DCAT datasets, in total consisting of more than 170 million \ac{RDF} triples, fetched from more than 80 data providers. Open Data is considered to be a key building block of Europe's data economy \cite{eu_open-data}, indicating the practical relevance of our work.

\subsection{Impact of Semantic Web Technologies}\label{sec:swtech}
The initially required development effort was higher and partly more challenging than with more traditional approaches. Some artifacts of the Semantic Web have not yet reached the required production readiness or caught up with latest progresses in software development. This increased integration effort and required some interim solutions for providing a production system. For instance, integrating synchronous third-party libraries into our asynchronous programming model. Particularly challenging was the adoption of a triplestore as primary database. The access is implemented on a very low level via SPARQL, since a mature object-relational mapping (ORM) tool does not exist. Most of the integrity and relationship management of the data is handled on application level and needed to be implemented there, since the triplestore, unlike relational databases, cannot handle constraints directly. In addition, the SPARQL endpoint should be openly available. This currently prevents the management of closed or draft data and will require a more elaborated approach. To the best of our knowledge no (free) production triplestore is available, supporting that kind of access control on the SPARQL endpoint. Furthermore, in the Open Data domain there is no suitable and mature method to present \ac{RDF} in a user interface. Hence, the transformation and processing of \ac{RDF} is still required before final presentation. Usually, this presentation is domain-depended and builds on custom implementations. We solved this by applying our search service for both, strong search capabilities and immediate presentation of the data in a user front-end. \\ 
However, the overall benefits outweigh the initial barriers and efforts. With our native application of the Semantic Web data model and its definite standards via a triplestore as principal data layer, we are much more able to harness the full potential of many Open Data specifications. This particularly concerns the required implementation of DCAT-AP. The direct reuse and linking to existing vocabularies or other resources enable a more expressive and explicit description of the data, e.g. for license, policy, and provenance information. In addition, this approach increases the machine-readability. The good supply of tools for working with \ac{RDF} simplifies the integration into third-party applications and creates new possibilities for browsing, processing, and understanding the data. Especially, the availability of tools for reasoning can support the creation of new insights and derived data. The native capabilities of \ac{RDF} to handle multiple languages support the cross-national aspect of Open Data. The application of \ac{SHACL} in connection with \ac{DQV} allowed us to generate and provide comprehensive quality information in a very effective fashion. In general, the strong liaison of the Semantic Web technologies facilitates a seamless integration of the data processing pipe.

\section{Conclusions and Outlook}
In this paper we have presented our scalable Open Data management platform Piveau. It provides functions for Open Data publication, quality assurance, and reuse, typically conducted by public administrations, research institutes and journalists. We applied a wide range of Semantic Web technologies and principles in our solution to overcome barriers and to address functional requirements of this domain. Although the Open Data community has always leveraged specifications of the Semantic Web, our work takes a previously untaken step by designing our platform around Semantic Web technologies from scratch. This allows for a much more efficient and immediate application of existing Open Data specifications. Hence, Piveau closes a gap between formal specifications and their utilization in production. We combined this with a new scalable architecture and an efficient development lice-cycle approach. Our orchestration approach enables a sustainable and flexible creation of Open Data platforms. Furthermore, it fosters the reuse of individual aspects of Piveau beyond the scope of Open Data. We have shown that our work can compete with existing Open Data solutions and exceed their features in several aspects. We have improved the generation and provision of quality information, enhanced the expressiveness of the metadata model and the support for multilingualism. As the core technology of the European Data Portal, Piveau promotes the Semantic Web as a highly relevant concept for Europe's data economy and has proven to be ready for production and reached a high degree of maturity. Finally, our work is a relevant contribution to the 5-star deployment scheme of Open Data, which supports the concept of Linked Open Data \cite{berners-lee_5stars}. The source code of Piveau can be found on GitLab.\footnote{\url{https://gitlab.com/piveau}} \\
In the next steps, Piveau will be extended with additional features. This includes support for user interaction, data enrichment, and data analysis. The support for further Semantic Web features is also planned, e.g. compliance with the LDP specifications and the extension beyond metadata to manage actual data as \ac{RDF}. Open research questions are the implementation of revision and access control on triplestore level, which cannot be satisfied yet on production-grade. In general, we aim to increase the overall readiness, broaden the target group beyond the Open Data community, and strengthen the meaning of Semantic Web technologies as core elements of data ecosystems.

\section*{Acknowledgments}
This work has been partially supported by the Federal Ministry of Education and Research of Germany (BMBF) under grant no. 16DII111 ("Deutsches Internet-Institut") and by the EU Horizon 2020 project "Reflow" under grant agreement no. 820937. The implementation and provision of the European Data Portal is funded by the European Commission under contracts DG CONNECT SMART 2014/1072 and SMART 2017/1123.

\newpage

%
%
%
%
\bibliographystyle{splncs04}
\bibliography{main}
\end{document}